\documentclass[letterpaper, 10 pt, conference]{ieeeconf}

\IEEEoverridecommandlockouts   
\usepackage{
  amsmath,
  amssymb,
  amsfonts,
  booktabs
}

\usepackage{amsthm}

\usepackage{bm}
\usepackage{graphicx}
\usepackage{cite}
\usepackage{tikz}
\usepackage{makecell}
\usetikzlibrary{arrows.meta}
\newtheorem{theorem}{Theorem}

\newtheorem{proposition}{Proposition}

\newtheorem{assumption}{Assumption}

\title{Geometry--Induced Contraction Degradation and Stabilization of Learning-Enabled Observers}

\author{Aditi Acharya, \textit{Member, IEEE} and Andrew Fleck, \textit{Member, IEEE}
\thanks{The authors acknowledge in-kind support from Purdue University.}
\thanks{© 2026 IEEE.  Personal use of this material is permitted.  Permission from IEEE must be obtained for all other uses, in any current or future media, including reprinting/republishing this material for advertising or promotional purposes, creating new collective works, for resale or redistribution to servers or lists, or reuse of any copyrighted component of this work in other works.}
}

\begin{document}
\maketitle

\begin{abstract}
Learned perception models increasingly serve as measurement maps within
nonlinear observers. Their state-dependent Jacobians introduce measurement geometry that
measurement-value accuracy does not constrain. Even when measurements
satisfy standard accuracy criteria, the corresponding one-step Euclidean
error operator can be noncontractive. Improving measurement accuracy or
retuning the correction strength may leave this observer-level contraction
failure unresolved. We derive the exact estimation-error operator and use it to determine whether
scalar-gain adjustment can restore contraction or the correction structure
must change. The analysis separates three distinct cases: a conservative
certificate may be inconclusive, the nominal gain may be exactly
noncontractive, or the correction geometry may lie beyond scalar-gain repair.
Convexity makes this recovery boundary exactly computable. When scalar
recovery remains possible, uncertainty-aware envelopes define a safe
projection that changes the nominal gain only when contraction and same-state
improvement are certified. Uniform local certification yields exponential
error decay and explicit disturbance-dependent bounds. The method uses local
composite-Jacobian information or a certified approximation and requires no
retraining. Scalar experiments demonstrate certified contraction recovery and
improved tracking when the geometry permits. Controlled real-image experiments
provide numerical evidence of scalar-unrecoverable contraction loss despite
accurate learned measurements.
\end{abstract}
\section{Introduction}
Accurate learned measurements do not guarantee strict Euclidean contraction
of the observer error dynamics. Measurement-value accuracy does not constrain
the Jacobian geometry that determines correction magnitude and
predictor--correction orientation in the exact estimation-error operator.
Under fixed gain, this geometry can make the nominal operator noncontractive
even when the measurement values are accurate. Improving measurement accuracy
or retuning the correction strength may leave this observer-level contraction
failure unresolved. A sensitivity-based upper bound provides an inexpensive
sufficient certificate, but its failure establishes neither exact
noncontraction nor instability. Jacobian sensitivity magnitude alone also
cannot determine whether scalar-gain adjustment restores contraction, since
the exact induced factor depends on predictor--correction orientation.
\begin{figure}[!t]
    \centering
    \resizebox{0.98\columnwidth}{!}{%
    \begin{tikzpicture}[
        >=Latex,
        line cap=round,
        line join=round,
        font=\footnotesize
    ]
        \definecolor{ubblue}{RGB}{30,92,168}
        \definecolor{lbgray}{RGB}{105,105,105}
        \definecolor{safegreen}{RGB}{34,139,94}
        \definecolor{thresholdpurple}{RGB}{116,72,145}

        \fill[safegreen!11]
            (2.10,0) rectangle (5.11,4.35);

        \draw[->,semithick]
            (0.42,0) -- (6.72,0)
            node[anchor=west] {$\alpha$};

        \draw[->,semithick]
            (0.42,0) -- (0.42,4.52)
            node[anchor=south west,align=left]
            {induced\\factor};

        \draw[
            dash pattern=on 5pt off 2.5pt,
            thresholdpurple
        ]
            (0.42,2.48) -- (6.55,2.48)
            node[pos=0,anchor=east]
            {$b_t$};

        \draw[lbgray,densely dashed,thick]
            (0.58,3.50)
            .. controls (1.55,2.56) and (2.43,1.10)
            .. (3.27,0.92)
            .. controls (4.15,0.83) and (5.20,1.90)
            .. (6.34,2.94);

        \draw[black,very thick]
            (0.58,3.84)
            .. controls (1.60,2.88) and (2.46,1.52)
            .. (3.29,1.30)
            .. controls (4.18,1.19) and (5.22,2.21)
            .. (6.34,3.27);

        \draw[ubblue,very thick]
            (0.58,4.17)
            .. controls (1.61,3.18) and (2.48,1.92)
            .. (3.30,1.67)
            .. controls (4.20,1.54) and (5.24,2.54)
            .. (6.34,3.62);

        \node[
            ubblue,
            fill=white,
            inner sep=1.2pt
        ] at (5.82,3.88)
            {$U_t(\alpha)$};

        \node[
            fill=white,
            inner sep=1.2pt
        ] at (5.74,3.02)
            {$\rho_t(\alpha)$};

        \node[
            lbgray,
            fill=white,
            inner sep=1.2pt
        ] at (5.69,2.11)
            {$L_t(\alpha)$};

        \node[
            safegreen!55!black,
            align=center
        ] at (3.60,4.02)
            {safely admissible\\set $\widehat{\mathcal S}_t(q,\Delta_\rho).$};

        \draw[densely dashed,black!65]
            (0.91,0) -- (0.91,3.55);

        \draw[densely dashed,safegreen!75!black]
            (2.10,0) -- (2.10,2.48);

        \fill
            (0.91,0) circle (1.5pt);

        \fill[safegreen!75!black]
            (2.10,0) circle (1.7pt);

        \node[anchor=north]
            at (0.91,-0.05)
            {$\alpha_0$};

        \node[
            safegreen!55!black,
            anchor=north
        ] at (2.10,-0.05)
            {$\alpha_t$};

        \draw[
            ->,
            semithick,
            safegreen!65!black
        ]
            (1.02,0.62) -- (1.98,0.62)
            node[midway,above]
            {project};

    \end{tikzpicture}%
    }
\caption{Robust safe projection for
$A_t(\alpha)=F_t-\alpha G_t$. The envelopes satisfy
$L_t(\alpha)\le \rho_t(\alpha)\le U_t(\alpha)$, and
$b_t:=\min\{q,L_t(\alpha_0)-\Delta_\rho\}$ is the effective
certification threshold. The shaded interval is
$\widehat{\mathcal S}_t(q,\Delta_\rho)$. When projection is required,
$\alpha_t$ is its point nearest $\alpha_0$. If the interval is empty, the
nominal gain is retained and certificate unavailability is reported. An
already-certified nominal gain is retained unchanged.}
    \label{fig:safe_projection}
\end{figure}
Learned measurement maps arise in hybrid learning/model-based estimators that
retain recursive structure while learning components that are difficult to
model analytically
\cite{Revach2022KalmanNet,Liu2024NeuralEKF,StateEstimationMLKF2023}.
Such components may define measurement interfaces or observer-coordinate maps
for high-dimensional inputs and partially known plant or measurement models
\cite{Liu2024NeuralEKF,Miao2023LearningRobustObservers,Guo2026KILOEKF}.
Viewing conditions, scale, occlusion, scattering, and distortion challenge
visual tracking
\cite{VisDrone,Panetta2021UOT,Alawode2022UTB180}. These variations can also
alter the local measurement Jacobian presented to the observer.
Learning-enabled observer designs include learned KKL or Luenberger
transformations and inverse reconstruction maps
\cite{PeralezNadri2021,Niazi2023LearningLuenberger,Tang2023LipschitzObserver}, with switching and nonautonomous extensions
\cite{PeralezNadri2024Switching,Woelk2026NeuralLuenberger}. Other work
analyzes neural approximation error \cite{LMIbasedneuralobserver2024}, learns
contraction-compatible corrections or metrics
\cite{Marani2025ContractingObserver,Marani2026RobustContraction,
TsukamotoChung2021NCM}, or jointly learns KKL transformations and contracting
latent dynamics \cite{Galimberti2026ContractingKKL}. Numerical KKL tuning and
switched high-gain observers address convergence speed and noise sensitivity
\cite{BuissonFenet2023GainTuning,AhrensKhalil2009SwitchedGain}.

Contraction theory provides differential conditions for incremental stability
and metric-observer design
\cite{LohmillerSlotine1996MetricObserver,LohmillerSlotine1998}, with
applications to extended Kalman filters and stochastic observers
\cite{BonnabelSlotine2015,DaniChungHutchinson2015}. Differential Lyapunov and
convex contraction methods support incremental analysis and nonlinear observer
synthesis
\cite{ForniSepulchre2014,Manchester2018ContractingObservers,
Yi2022ReducedOrder}, including privacy-preserving observer design
\cite{LeNy2020DifferentiallyPrivate}. Unlike these design-oriented approaches, our analysis fixes the learned
measurement map and correction matrix, retains the Euclidean induced norm, and characterizes exact contraction recoverability over the admissible scalar-gain family.

\textbf{Contributions.}
\begin{itemize}
\item \emph{Geometry-induced contraction degradation.}
We derive the exact Euclidean error operator
$A_t(\alpha)=F_t-\alpha G_t$, where
$G_t=\widetilde K\bar J_{h,t}$. Contraction depends on the correction
magnitude $G_t^\top G_t$ and predictor--correction orientation
$F_t^\top G_t+G_t^\top F_t$, neither of which is constrained by
measurement-value accuracy. An accurate learned map may still yield
$\rho_t(\alpha_0)\ge1$. This separates failure of a conservative sensitivity
certificate from exact nominal noncontraction.
\item \emph{Exact recoverability, structural impossibility, and certified
intervention.}
We establish convexity of
$\rho_t(\alpha)=\|A_t(\alpha)\|_2$, characterize its target sublevel sets by
an exact LMI, and obtain a necessary-and-sufficient scalar-recoverability
test. An uncorrected-direction bound certifies cases in which no admissible scalar gain recovers strict contraction. Uncertainty-aware envelopes define a nearest-gain
projection that changes the nominal gain only when the contraction target and
exact same-state improvement are certified.
\item \emph{Robust convergence and empirical recoverability limits.}
Uniform certification yields exponential error decay in the unforced case and
explicit disturbance-dependent bounds. Scalar experiments recover contraction and reduce estimation error
when recovery is feasible. Controlled real-image experiments provide numerical
evidence of scalar-unrecoverable contraction loss despite accurate
measurements.
\end{itemize}
The framework is independent of measurement-map architecture and training and
does not modify the learned parameters. It requires local model evaluations
and a composite Jacobian or certified approximation whose error enters the
mismatch envelope.
\section{Problem Formulation and Exact Error Dynamics}
\label{sec:theory}
Consider the discrete-time nonlinear system and learned measurement model
\begin{equation}
x_{t+1}=f(x_t)+w_t,
\qquad
y_t=h_\phi(x_t)+v_t,
\label{eq:plant_measurement}
\end{equation}
and the scalar-gain observer
\begin{equation}
\hat x_{t+1}
=f(\hat x_t)+\alpha_t\widetilde K
\bigl(y_t-h_\phi(\hat x_t)\bigr),
\qquad \alpha_t\in[0,\alpha_{\max}],
\label{eq:observer}
\end{equation}
where $x_t\in\mathbb R^n$, $y_t\in\mathbb R^m$, and $w_t$ and $v_t$
denote process and measurement disturbances.  The differentiable map
$h_\phi:\mathbb R^n\to\mathbb R^m$ is frozen after training.  

The matrix
$\widetilde K\in\mathbb R^{n\times m}$ is the fixed correction matrix,
$\alpha_{\max}>0$, and $\alpha_0\in[0,\alpha_{\max}]$ is the nominal fixed
gain. 
Define $e_t:=x_t-\hat x_t$.  Throughout, $\|\cdot\|_2$ denotes the
Euclidean vector norm and its induced spectral matrix norm.

Write the predictor and measurement Jacobians as
\begin{equation}
J_f(x):=\frac{\partial f(x)}{\partial x},
\qquad
J_h(x):=\frac{\partial h_\phi(x)}{\partial x}.
\end{equation}

\begin{assumption}[Local regularity]
\label{ass:regularity}
On a region $\mathcal X\subset\mathbb R^n$, the maps $f$ and $h_\phi$ are
continuously differentiable and
\begin{align}
\|J_f(x)\|_2&\le L_f,\label{eq:Jf_bound}\\
\|J_f(x)-J_f(z)\|_2&\le L_{J_f}\|x-z\|_2,\label{eq:Jf_lipschitz}\\
\|J_h(x)-J_h(z)\|_2&\le L_{J_h}\|x-z\|_2
\label{eq:Jh_lipschitz}
\end{align}
for all $x,z\in\mathcal X$.
\end{assumption}

Whenever the segment $\{\hat x_t+s e_t:s\in[0,1]\}$ lies in $\mathcal X$,
define
\begin{equation}
F_t:=\int_0^1J_f(\hat x_t+s e_t)\,ds,
\qquad
G_t:=\widetilde K\int_0^1J_h(\hat x_t+s e_t)\,ds.
\label{eq:averaged_operators}
\end{equation}
The integral mean-value identity gives
\begin{equation}
f(x_t)-f(\hat x_t)=F_te_t,
\qquad
\widetilde K\bigl(h_\phi(x_t)-h_\phi(\hat x_t)\bigr)=G_te_t,
\end{equation}
and therefore the exact recursion
\begin{equation}
\begin{aligned}
e_{t+1}
&=A_t(\alpha_t)e_t+d_t(\alpha_t),\\
A_t(\alpha)
&:=F_t-\alpha G_t,
\qquad
d_t(\alpha):=w_t-\alpha\widetilde K v_t.
\end{aligned}
\label{eq:error_recursion}
\end{equation}
If $\|w_t\|_2\le\bar w$ and $\|v_t\|_2\le\bar v$, then
\begin{equation}
\begin{aligned}
\|d_t(\alpha)\|_2
&\le d_{\max}(\alpha),\\
d_{\max}(\alpha)
&:=\bar w+\alpha\|\widetilde K\|_2\bar v.
\end{aligned}
\label{eq:gain_dependent_disturbance}
\end{equation}
Equation~\eqref{eq:error_recursion} exposes the central failure mechanism:
prediction accuracy constrains measurement-value error, while the learned
Jacobian enters $G_t$ and shapes how the correction acts across error
directions. An accurate learned map can therefore make the nominal operator
$F_t-\alpha_0G_t$ noncontractive. Varying $\alpha$ moves the exact error operator only along the affine family
$F_t-\alpha G_t$. Its restriction to $\ker(G_t)$ is independent of
$\alpha$:
$A_t(\alpha)z=F_tz$ for every $z\in\ker(G_t)$.

\subsection{Exact Geometry-Dependent Contraction and Recoverability}
We characterize the exact scalar-recovery boundary, distinguishing nominal
noncontraction that scalar-gain adjustment can repair from correction geometry
for which strict contraction is unattainable over the admissible gain family.
For fixed $x_t$ and $\hat x_t$, $\rho_t(\alpha)$ is the worst-case one-step
amplification of $A_t(\alpha)$ over all nonzero Euclidean directions. For the
realized error $e_t\neq0$,
\[
\frac{\|A_t(\alpha)e_t\|_2}{\|e_t\|_2}
\]
is the amplification along that direction. Hence
$\rho_t(\alpha)\ge1$ rules out strict contraction in every direction but does
not imply growth of the current error. In contrast,
$\rho_t(\alpha)<1$ holds if and only if $A_t(\alpha)$ contracts every
Euclidean direction and guarantees strict one-step error decrease in the
unforced recursion.
This distinction defines a hierarchy of failure modes. Failure of a
sufficient bound may reflect conservatism, whereas
$\rho_t(\alpha_0)\ge1$ establishes exact noncontraction of the nominal
operator. Scalar-family unrecoverability is stronger: no admissible gain
makes the exact operator strictly contractive. For a requested factor $q$,
the exact feasible set contains all gains satisfying
$\rho_t(\alpha)\le q$, and $\gamma_t^\star$ is the smallest exact factor
attainable over the admissible scalar family.
\begin{theorem}[Exact scalar-gain recoverability]
\label{thm:recoverability}
For $q\ge0$, let
\begin{equation}
\begin{aligned}
\mathcal A_t^{\rm ex}(q)
&:=\bigl\{\alpha\in[0,\alpha_{\max}]
\mid \rho_t(\alpha)\le q\bigr\},\\
\gamma_t^\star
&:=\min_{\alpha\in[0,\alpha_{\max}]}\rho_t(\alpha).
\end{aligned}
\label{eq:exact_set_gamma}
\end{equation}
The map $\rho_t$ is continuous and convex.  Consequently, its minimizer set
and every nonempty set $\mathcal A_t^{\rm ex}(q)$ are closed intervals
(possibly singletons), and
\begin{equation}
\mathcal A_t^{\rm ex}(q)\neq\varnothing
\quad\Longleftrightarrow\quad
\gamma_t^\star\le q.
\label{eq:target_recoverability}
\end{equation}
Moreover,
\begin{equation}
\alpha\in\mathcal A_t^{\rm ex}(q)
\quad\Longleftrightarrow\quad
\begin{bmatrix}
qI & F_t-\alpha G_t\\
(F_t-\alpha G_t)^\top & qI
\end{bmatrix}\succeq0.
\label{eq:exact_lmi}
\end{equation}
Thus strict Euclidean contraction is recoverable in the allowed scalar family if and only if $\gamma_t^\star<1$; it is unrecoverable if
$\gamma_t^\star\ge1$.  The case $\gamma_t^\star=1$ reaches the boundary but does not recover strict contraction.
\end{theorem}
\begin{proof}
The spectral norm is continuous and convex, so $\rho_t$ is continuous and
convex in $\alpha$. Continuity on the compact admissible interval guarantees
a minimizer, while convexity gives the stated closed-interval structure and
\eqref{eq:target_recoverability}. For $q>0$, \eqref{eq:exact_lmi} follows by
the Schur complement. For $q=0$, both sides require
$F_t-\alpha G_t=0$.
\end{proof}
The value $\gamma_t^\star$ therefore converts a failed nominal test into a
structural diagnosis.  If $\rho_t(\alpha_0)\ge1$ but
$\gamma_t^\star<1$, contraction has been lost at the nominal gain but remains
recoverable by scalar rescaling.  If $\gamma_t^\star\ge1$, normalization has
not just selected the wrong gain: \textit{no gain in the admissible scalar family
recovers strict Euclidean contraction} for the fixed values of $x_t$ and
$\hat x_t$.
The exact contraction condition also has the matrix form
\begin{equation}
\rho_t(\alpha)<1
\quad\Longleftrightarrow\quad
I-A_t(\alpha)^\top A_t(\alpha)\succ0.
\label{eq:exact_pd_condition}
\end{equation}
Expanding its left-hand side gives
\begin{equation}
\begin{aligned}
I-A_t(\alpha)^\top A_t(\alpha)
={}&I-F_t^\top F_t\\
&+\alpha\bigl(F_t^\top G_t+G_t^\top F_t\bigr)\\
&-\alpha^2G_t^\top G_t.
\end{aligned}
\label{eq:orientation_expansion}
\end{equation}
This expansion shows why sensitivity magnitude alone is insufficient:
$G_t^\top G_t$
records correction magnitude, while
$F_t^\top G_t+G_t^\top F_t$ records predictor--correction orientation.

\begin{proposition}[Dominant-direction alignment]
\label{prop:alignment}
Fix $t$ and suppose that the dominant singular value $\rho_t(\alpha)>0$ of
$A_t(\alpha)$ is simple.  Let $\ell_t$ and $r_t$ be corresponding unit left and
right singular vectors.  Then
\begin{equation}
\frac{d\rho_t(\alpha)}{d\alpha}=-\ell_t^\top G_tr_t.
\label{eq:alignment_derivative}
\end{equation}
Consequently, increasing $\alpha$ locally improves the exact factor when
$\ell_t^\top G_tr_t>0$ and degrades it when $\ell_t^\top G_tr_t<0$.
\end{proposition}

\begin{proof}
The standard perturbation formula for a simple singular value gives
$d\rho_t/d\alpha=\ell_t^\top(dA_t/d\alpha)r_t$.  Since
$dA_t/d\alpha=-G_t$, \eqref{eq:alignment_derivative} follows.
\end{proof}

At a repeated dominant singular value, local gain dependence is governed by
the spectral-norm subdifferential rather than a single singular-vector pair.
Thus $\|G_t\|_2$ alone cannot determine improvement. Moreover, for any
$z\in\ker(G_t)$, $A_t(\alpha)z=F_tz$ for all $\alpha$, so scalar gain cannot
alter error propagation along $z$.
\begin{theorem}[Uncorrected-direction obstruction]
\label{thm:nullspace_obstruction}
Let $P_{L,t}$ and $P_{R,t}$ be the orthogonal projectors onto
$\ker(G_t^\top)$ and $\ker(G_t)$, respectively.  Then
\begin{equation}
\gamma_t^\star\ge
\max\{\|P_{L,t}F_t\|_2,\|F_tP_{R,t}\|_2\}.
\label{eq:nullspace_lower_bound}
\end{equation}
If the right-hand side is at least one, strict Euclidean contraction is
unrecoverable within the scalar family.
\end{theorem}

\begin{proof}
Since $P_{L,t}G_t=0$ and $G_tP_{R,t}=0$,
\begin{align}
\|F_t-\alpha G_t\|_2&\ge\|P_{L,t}F_t\|_2,\\
\|F_t-\alpha G_t\|_2&\ge\|F_tP_{R,t}\|_2.
\end{align}
Minimizing over $\alpha$ proves the claim.
\end{proof}
This obstruction is specific to the Euclidean metric, the prescribed
direction $\widetilde K$, and the scalar gain family.  A lower bound below one
does not establish recoverability, and the result does not rule out another
metric, matrix-valued gain, or observer architecture.

\subsection{The Conservative Sensitivity Certificate}
The sensitivity condition provides a computationally inexpensive sufficient
certificate but does not characterize the exact contraction boundary. At the nominal
gain, the triangle inequality yields the oracle bound
\begin{equation}
\rho_t(\alpha_0)\le L_f+\alpha_0s_t,
\qquad s_t:=\|G_t\|_2.
\label{eq:fixed_triangle_bound}
\end{equation}
If $L_f<1$ and $\alpha_0>0$, then
$s_t<(1-L_f)/\alpha_0$ certifies contraction.  Crossing this sensitivity
threshold loses only the sufficient bound; the exact operator may remain
contractive through favorable orientation and cancellation.  Thus
$L_f+\alpha_0s_t=1$ is a certificate-loss boundary, whereas
$\rho_t(\alpha_0)=1$ is the exact boundary.  The sensitivity-only certificate
ignores orientation, cancellation, and rank-deficient uncorrected directions,
and it does not imply that $\rho_t$ is monotone in $s_t$.

\subsection{Implementable robust envelopes}
The exact quantities $F_t$, $G_t$, and $\gamma_t^\star$ characterize the
scalar-recovery boundary, but their evaluation requires the unknown segment
$\{\hat x_t+s(x_t-\hat x_t):s\in[0,1]\}$. Thus, they are unavailable from
the observer information at time $t$. The implementable rule uses Jacobian
evaluations at $\hat x_t$ and explicit bounds on the segment mismatch.
Define
\begin{equation}
\widehat F_t:=J_f(\hat x_t),
\qquad
\widehat G_t:=\widetilde K\widehat J_{h,t},
\label{eq:estimated_operators}
\end{equation}
where
\begin{equation}
\|\widehat J_{h,t}-J_h(\hat x_t)\|_2\le\delta_J.
\label{eq:jacobian_error}
\end{equation}
On the local tube $\|e_t\|_2\le r$, Assumption~\ref{ass:regularity} gives
\begin{align}
\|F_t-\widehat F_t\|_2
&\le c_f(r):=\frac{L_{J_f}r}{2},\label{eq:F_mismatch}\\
\|G_t-\widehat G_t\|_2
&\le c_h(r):=\|\widetilde K\|_2
\left(\delta_J+\frac{L_{J_h}r}{2}\right).
\label{eq:G_mismatch}
\end{align}
Hence, for $\alpha\in[0,\alpha_{\max}]$,
\begin{align}
\widehat\rho_t(\alpha)&:=\|\widehat F_t-\alpha\widehat G_t\|_2,\\
\eta_t(\alpha)&:=c_f(r)+\alpha c_h(r),\\
U_t(\alpha)&:=\widehat\rho_t(\alpha)+\eta_t(\alpha),\label{eq:upper_envelope}\\
L_t(\alpha)&:=\max\{0,\widehat\rho_t(\alpha)-\eta_t(\alpha)\}
\label{eq:lower_envelope}
\end{align}
satisfy
\begin{equation}
L_t(\alpha)\le\rho_t(\alpha)\le U_t(\alpha).
\label{eq:envelope_bound}
\end{equation}
The triangle and reverse-triangle inequalities applied to
$(F_t-\alpha G_t)-(\widehat F_t-\alpha\widehat G_t)$ give
\eqref{eq:envelope_bound}. Hence
$U_t(\alpha)\le L_t(\alpha_0)-\Delta_\rho$ certifies
$\rho_t(\alpha)\le\rho_t(\alpha_0)-\Delta_\rho$ at the fixed state pair.
The norm of an affine matrix function is continuous and convex, so
$\widehat\rho_t$ and $U_t=\widehat\rho_t+\eta_t$ are continuous and convex on
$[0,\alpha_{\max}]$.
For $q\in[0,1)$, define the robust inner approximation
\begin{equation}
\widehat{\mathcal A}_t(q)
:=\{\alpha\in[0,\alpha_{\max}]:U_t(\alpha)\le q\}.
\label{eq:robust_set}
\end{equation}
Consequently, this set is empty or a closed
interval, and
\begin{equation}
\widehat{\mathcal A}_t(q)\subseteq\mathcal A_t^{\rm ex}(q).
\label{eq:inner_inclusion}
\end{equation}
Its constraint is equivalently
\begin{equation}
\begin{bmatrix}
(q-\eta_t(\alpha))I & \widehat F_t-\alpha\widehat G_t\\
(\widehat F_t-\alpha\widehat G_t)^\top &
(q-\eta_t(\alpha))I
\end{bmatrix}\succeq0,
\label{eq:robust_lmi}
\end{equation}
which also enforces $q-\eta_t(\alpha)\ge0$.  An empty robust set establishes only failure of this certificate, not exact
noncontraction or scalar-family infeasibility.

\subsection{Safe representation-aware projection}
Fix a desired reserve $\mu_\star\in(0,1)$ and set $q:=1-\mu_\star$.
Projection onto \eqref{eq:robust_set} enforces the target but does not alone
prove that changing the gain improves upon the exact nominal operator.  For a
requested same-state improvement $\Delta_\rho\ge0$, define
\begin{equation}
\begin{aligned}
\widehat{\mathcal S}_t(q,\Delta_\rho)
&:=\widehat{\mathcal A}_t(q)\cap\\
&\quad\Bigl\{\alpha\in[0,\alpha_{\max}]\,\Bigm|\\
&\qquad U_t(\alpha)
\le L_t(\alpha_0)-\Delta_\rho\Bigr\}.
\end{aligned}
\label{eq:safe_set}
\end{equation}
This set is empty or a closed interval.  Select
\begin{equation}
\alpha_t=
\begin{cases}
\displaystyle
\operatorname*{arg\,min}_{\alpha\in
\widehat{\mathcal S}_t(q,\Delta_\rho)}
\tfrac12(\alpha-\alpha_0)^2,
&
\substack{
U_t(\alpha_0)>q,\\
\widehat{\mathcal S}_t(q,\Delta_\rho)\neq\varnothing,
}\\[4mm]
\alpha_0,
&\text{otherwise}.
\end{cases}
\label{eq:safe_projection}
\end{equation}
If $U_t(\alpha_0)>q$ and
$\widehat{\mathcal S}_t(q,\Delta_\rho)=\varnothing$, the rule retains
$\alpha_0$ and reports certificate unavailability. The projection is unique and may increase or decrease the gain. If the
nominal gain satisfies $U_t(\alpha_0)\le q$, it is retained. Otherwise, the
gain changes only when the envelopes certify
\[
\rho_t(\alpha)\le q,
\qquad
\rho_t(\alpha)\le\rho_t(\alpha_0)-\Delta_\rho
\]
at the fixed state pair. If no such gain is certified, the nominal gain is
retained and certificate unavailability is reported.
\begin{theorem}[Certified same-state dominance]
\label{thm:nondegradation}
Under Assumption~\ref{ass:regularity}, suppose
\eqref{eq:jacobian_error} holds, $\|e_t\|_2\le r$, and
$\{\hat x_t+s e_t:s\in[0,1]\}\subseteq\mathcal X$. Rule
\eqref{eq:safe_projection} satisfies
\begin{equation}
\rho_t(\alpha_t)\le\rho_t(\alpha_0).
\label{eq:exact_nondegradation}
\end{equation}
If the projection branch is used, then
\begin{equation}
\rho_t(\alpha_t)
\le\rho_t(\alpha_0)-\Delta_\rho,
\qquad
\rho_t(\alpha_t)\le1-\mu_\star.
\label{eq:strict_improvement_target}
\end{equation}
Thus every intervention with $\Delta_\rho>0$ strictly improves the exact
one-step induced factor by at least $\Delta_\rho$ while enforcing the desired
reserve, with $F_t$ and $G_t$ fixed at time $t$.
\end{theorem}
\begin{proof}
On either no-intervention branch, $\alpha_t=\alpha_0$ and
\eqref{eq:exact_nondegradation} is equality.  On the projection branch,
\eqref{eq:envelope_bound} and feasibility give
\begin{equation}
\rho_t(\alpha_t)
\le U_t(\alpha_t)
\le L_t(\alpha_0)-\Delta_\rho
\le\rho_t(\alpha_0)-\Delta_\rho,
\end{equation}
and $\rho_t(\alpha_t)\le U_t(\alpha_t)\le1-\mu_\star$.
\end{proof}
The theorem compares the norms of $F_t-\alpha_tG_t$ and
$F_t-\alpha_0G_t$ for fixed $F_t$ and $G_t$. It does not order the
subsequent estimation errors after the two gain choices generate different
values of $\hat x_{t+1}$. Induced-norm dominance also does not imply the
directional ordering
\begin{equation}
\|A_t(\alpha_t)e_t\|_2
\le \|A_t(\alpha_0)e_t\|_2,
\label{eq:no_directional_claim}
\end{equation}
nor does it guarantee pointwise error reduction or lower RMSE over the
resulting estimate sequences. 

When the nominal gain is not already certified, three cases must be
distinguished. If $\gamma_t^\star>q$, the admissible scalar family cannot
attain the requested target. If $\gamma_t^\star\le q$ but
$\widehat{\mathcal A}_t(q)=\varnothing$, the target is exactly feasible but
not certified under the stated mismatch bounds. If
$\widehat{\mathcal A}_t(q)\neq\varnothing$ but
$\widehat{\mathcal S}_t(q,\Delta_\rho)=\varnothing$, the target is certified
for some gain, but the envelopes do not certify the requested same-state
improvement. Only $\gamma_t^\star\ge1$ establishes strict-contraction
unrecoverability within the scalar family.

\subsection{Disturbance robustness}

The certified same-state comparison in Theorem~\ref{thm:nondegradation} does
not by itself yield a multi-step error bound.  Such a bound requires
$U_t(\alpha_t)\le1-\mu_\star$ at every time step for which
$\|e_t\|_2\le r$.  The disturbance injection must also be small enough to
keep $\|e_t\|_2$ within that tube.

\begin{theorem}[Local robust convergence]
\label{thm:disturbance}
Suppose the local assumptions hold whenever $\|e_t\|_2\le r$ and the applied
gain satisfies
\begin{equation}
U_t(\alpha_t)\le1-\mu_\star<1
\label{eq:uniform_certified_target}
\end{equation}
at every step in that tube.  Let
\begin{equation}
\bar\alpha:=\sup_t\alpha_t\le\alpha_{\max},
\qquad
\bar d:=\bar w+\bar\alpha\|\widetilde K\|_2\bar v.
\label{eq:dbar_projection}
\end{equation}
If $\|e_0\|_2\le r$ and $\bar d\le\mu_\star r$, then the tube is forward
invariant and
\begin{equation}
\begin{aligned}
\|e_t\|_2
&\le (1-\mu_\star)^t\|e_0\|_2\\
&\quad
+\frac{1-(1-\mu_\star)^t}{\mu_\star}\bar d,\\
\limsup_{t\to\infty}\|e_t\|_2
&\le\frac{\bar d}{\mu_\star}.
\end{aligned}
\label{eq:disturbance_result}
\end{equation}
\end{theorem}

\begin{proof}
Within the tube,
$\|e_{t+1}\|_2\le(1-\mu_\star)\|e_t\|_2+\bar d$.
The condition $\bar d\le\mu_\star r$ proves forward invariance by induction;
iterating the scalar recursion gives \eqref{eq:disturbance_result}.
\end{proof}
The factor $1/\mu_\star$ is the certified disturbance-amplification factor,
and $\bar d/\mu_\star$ is the ultimate error radius. 
If the
certificate-unavailable fallback occurs at any step with
$U_t(\alpha_0)>1-\mu_\star$, the multi-step guarantee no longer
follows, even as retaining $\alpha_0$ leaves $\rho_t(\alpha_0)$ unchanged
at that time step. Recovering contraction does not by itself control the effect of measurement
noise on the error recursion. Additionally, if the safe projection increases $\alpha_t$, it increases the bound
$\alpha_t\|\widetilde K\|_2\bar v$ on the measurement-noise term. 
To certify a
requested ultimate radius $R_\star\in(0,r]$, we augment
\eqref{eq:safe_set} with the convex constraint
\begin{equation}
U_t(\alpha)+\frac{d_{\max}(\alpha)}{R_\star}\le1.
\label{eq:radius_constraint}
\end{equation}
For this variant, $\alpha_0$ is retained as certified only when
$U_t(\alpha_0)\le q$ and it satisfies \eqref{eq:radius_constraint}.
Otherwise, the gain is projected onto the safe set intersected with this
constraint. If that intersection is empty, the rule retains $\alpha_0$ and
reports certificate unavailability. When the applied gain satisfies both
constraints at every step,
$d_{\max}(\alpha_t)\le(1-U_t(\alpha_t))R_\star$, and
\eqref{eq:uniform_certified_target} yields
\begin{equation}
[\|e_{t+1}\|_2-R_\star]_+
\le(1-\mu_\star)[\|e_t\|_2-R_\star]_+.
\end{equation}
The radius-$r$ tube is then invariant and
$\limsup_t\|e_t\|_2\le R_\star$. A fallback caused by failure of the radius
constraint removes this ultimate-radius guarantee. If
$U_t(\alpha_0)>1-\mu_\star$ also holds, the uniform contraction guarantee no
longer follows.

\subsection{Low-Cost Sensitivity-Cap Alternative}
The safe projection accounts for predictor--correction orientation and
Jacobian uncertainty but entails solving a state-dependent scalar convex
program.  For settings with stringent online computational constraints, we
therefore consider a closed-form sensitivity cap.
When only a sensitivity-magnitude rule is desired and $\alpha_0>0$, define
\begin{equation}
\widehat s_t:=\|\widehat G_t\|_2,
\qquad
\alpha_t^{\rm cap}
:=\min\left\{\alpha_0,
\frac{\beta}{\widehat s_t+\varepsilon}\right\},
\qquad \beta,\varepsilon>0.
\label{eq:capped_gain}
\end{equation}
Then $0<\alpha_t^{\rm cap}\le\alpha_0$ and
$\alpha_t^{\rm cap}\widehat s_t\le\beta$.  On $\|e_t\|_2\le r$,
\begin{equation}
\rho_t(\alpha_t^{\rm cap})
\le L_f+\beta+\alpha_0c_h(r)
=:q_{\rm cap}.
\label{eq:q_cap}
\end{equation}
Hence $q_{\rm cap}<1$ gives a uniform local sufficient certificate.  The cap
also improves the robust triangle certificate
\begin{equation}
B_t^{\rm tri}(\alpha)
:=\|\widehat F_t\|_2+c_f(r)
+\alpha(\widehat s_t+c_h(r))
\end{equation}
by
\begin{equation}
\begin{aligned}
&B_t^{\rm tri}(\alpha_0)
-B_t^{\rm tri}(\alpha_t^{\rm cap})\\
&\quad
=(\alpha_0-\alpha_t^{\rm cap})
  \bigl(\widehat s_t+c_h(r)\bigr)\\
&\quad
\ge \bigl[\alpha_0\widehat s_t-\beta\bigr]_+.
\end{aligned}
\label{eq:cap_margin_improvement}
\end{equation}
This guarantee is deliberately certificate-level: the cap improves the robust
triangle bound but need not decrease the exact factor $\rho_t$.  However, the cap is limited to gain-reduction and cannot use favorable alignment
for which stronger correction would improve contraction.  Reducing the gain
may instead increase $\rho_t$ when the correction offsets an expansive
predictor.  Accordingly, the cap is a low-cost alternative that provides
only sufficient certificate-level guarantees.  By contrast, the safe projection in \eqref{eq:safe_projection} uses the
estimated matrices $\widehat F_t$ and $\widehat G_t$ together with the
mismatch envelopes to certify the contraction target and same-state dominance
before changing the gain.

\subsection{Scope and implementation}
Measurement-prediction accuracy does not determine the local error dynamics
because the exact operator depends on the learned map's Jacobian. We
distinguish loss of the sensitivity certificate, exact noncontraction at the
nominal gain, and unrecoverability within the admissible scalar family. The
nominal factor $\rho_t(\alpha_0)$ tests the selected gain,
$\gamma_t^\star$ characterizes the best attainable scalar-gain factor, and
the robust envelopes certify contraction and same-state dominance online.
The rule retains an already-certified nominal gain, projects to the nearest
safe gain when the requested target and same-state dominance over
$F_t-\alpha_0G_t$ are certified with $F_t$ and $G_t$ fixed, and otherwise
retains the nominal gain while reporting certificate unavailability.
Multi-step disturbance guarantees require uniform
certification at every time step.
Implementation uses $\widehat F_t$, $\widehat G_t$, and explicit mismatch
bounds. If $h_\phi=p_\phi\circ g$, then $J_h$ is the composite Jacobian, and
all derivative-chain error must be included in $\delta_J$. Thus, ``black
box'' denotes a frozen differentiable interface with Jacobian access or a
certified approximation. Scalar-family unrecoverability is confined to the chosen metric and gain
family and does not preclude recovery through richer gains, alternative
metrics, or different observer architectures.

\section{Numerical Experiments}
\label{sec:numerical}
The scalar experiment provides the constructive case in which the nominal
operator loses contraction but the safe projection restores the requested
contraction target and disturbance bound. Section~\ref{sec:real_image}
provides the complementary numerical realization of scalar-unrecoverable
contraction loss.
\subsection{Certified Scalar Recovery}
\label{subsec:scalar_recovery}
Consider the scalar plant and nonlinear measurement model
\begin{equation}
x_{t+1}=0.5x_t+w_t,
\qquad
y_t=h_\kappa(x_t)+v_t,
\label{eq:scalar_experiment_model}
\end{equation}
where
\begin{equation}
h_\kappa(x)=x+\kappa x e^{-x^2},
\qquad
J_{h_\kappa}(x)
=1+\kappa e^{-x^2}(1-2x^2).
\label{eq:scalar_measurement_map}
\end{equation}

The observer uses $\widetilde K=1$, $\alpha_0=0.25$,
$\alpha\in[0,1]$, $r=0.06$, $q=0.45$ ($\mu_\star=0.55$), and
$\Delta_\rho=0$. The cap uses $\beta=0.30$ and $\varepsilon=10^{-3}$.
We use $L_{J_h}=1.95178307277\,\kappa$, $c_f=0$, and
$c_h=\delta_J+L_{J_h}r/2$. The perturbed Jacobian estimate is
$\widehat J_t^h=J_{h_\kappa}(\hat x_t)+\delta_J\xi_t$, where
$\xi_t\sim\mathcal U[-1,1]$. The same exogenous samples are used across
methods. The deterministic sweep uses $\kappa\in\{0,0.1,\ldots,6\}$ with no Jacobian
error and $(x_0,\hat x_0)=(0.25,0.20)$. In the stochastic arms,
$x_0\sim\mathcal U[-0.5,0.5]$, $e_0\sim\mathcal U[-0.06,0.06]$, and
$\hat x_0=x_0-e_0$. These arms use
$\kappa\in\{0.2,1,1.8,5,6\}$ and
$\delta_J\in\{0,0.02,0.05,0.10,0.20\}$, with $N=200$ paired 100-step trials
per condition. The bounded disturbances satisfy
$|w_t|\le5\times10^{-4}$ and $|v_t|\le5\times10^{-3}$. The Gaussian stress
test uses $\sigma_w=\sigma_v=10^{-2}$ and is not assigned a deterministic
disturbance certificate.

\subsubsection{Operator and Certificate Validation}
Across all 111 conditions, no certificate-unavailable fallback occurred.
Every selected gain satisfied $\rho_t(\alpha_t)\le q$ and same-state
nondegradation to a numerical tolerance of $10^{-12}$. The maximum nominal
factors are $1.244$, $1.250$, and $1.250$ in the deterministic,
bounded-noise, and Gaussian arms. The safe-projection maxima are $0.413$,
$0.439$, and $0.443$, all below $q=0.45$. Each arm contains noncontractive
nominal steps, while the safe projection satisfies the requested reserve at
every evaluated step.
\begin{figure}
    \centering
    \includegraphics[width=0.98\linewidth]
    {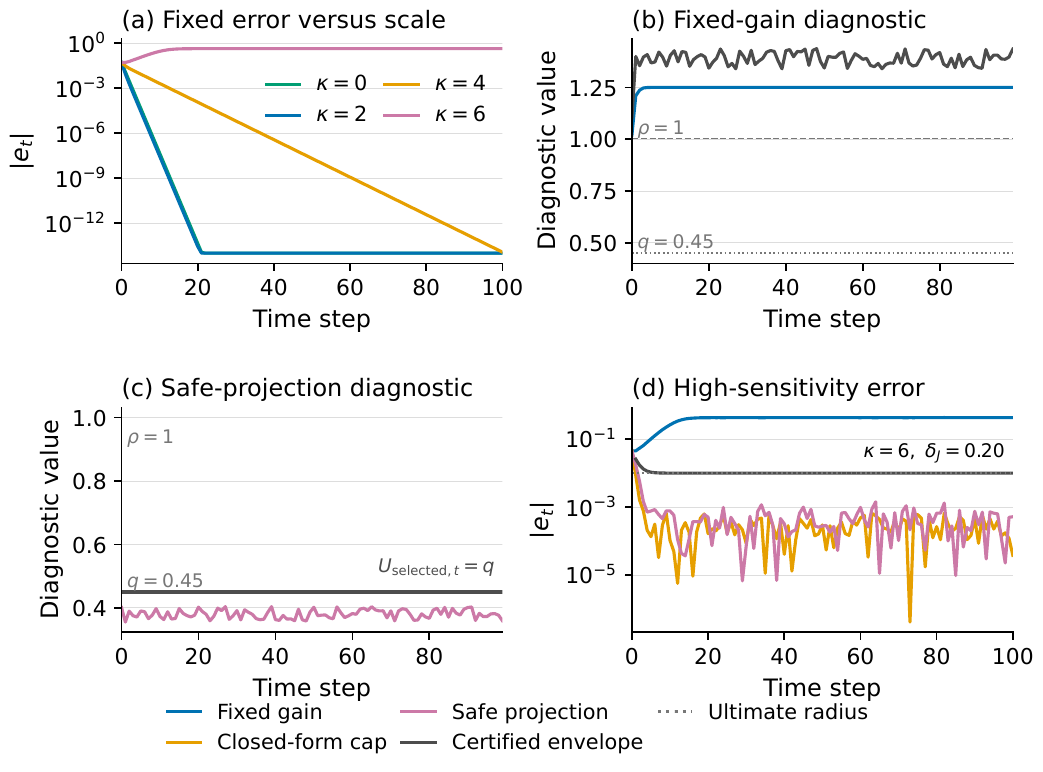}
    \caption{Certified scalar recovery under increasing measurement
    sensitivity. (a) Fixed-gain convergence is lost in the high-sensitivity
    regime at $\kappa=6$. (b) At $\kappa=6$, the exact fixed-gain factor and
    its robust upper bound exceed unity, certifying nominal noncontraction.
    (c) The safe projection satisfies
    $U_{\mathrm{selected},t}=q=0.45$, and the exact factor remains below $q$.
    (d) Under bounded disturbances, the safe projection and cap remain below
    the certified envelope approaching the ultimate radius $0.01$, while the
    fixed-gain error grows.}
    \label{fig:scalar_recovery}
\end{figure}

\subsubsection{Trajectory Performance and Disturbance Robustness} 
\begin{table}[t]
\centering
\caption{Scalar full-horizon RMSE over the evaluated conditions.}
\label{tab:scalar_rmse}
\setlength{\tabcolsep}{3.5pt}
\begin{tabular}{lccc}
\toprule
Arm & Fixed & Cap & Safe projection \\
\midrule
Deterministic & $1.1987\!\times\!10^{-1}$ &
$1.0629\!\times\!10^{-3}$ & $1.6219\!\times\!10^{-3}$ \\
Bounded & $1.7724\!\times\!10^{-1}$ &
$8.8519\!\times\!10^{-4}$ & $1.2057\!\times\!10^{-3}$ \\
Gaussian & $1.7905\!\times\!10^{-1}$ &
$1.0340\!\times\!10^{-2}$ & $1.0710\!\times\!10^{-2}$ \\
\bottomrule
\end{tabular}
\end{table}

The safe projection reduces pooled RMSE relative to fixed gain by $98.65\%$,
$99.32\%$, and $94.02\%$ across the three arms and worsens no evaluated
condition. These trajectory improvements are empirical and do not follow from
the same-state theorem. The cap has lower pooled RMSE but worsens 20 of 61
deterministic, 6 of 25 bounded-noise, and 9 of 25 Gaussian conditions. Additionally, it has
no exact same-state dominance guarantee. For the bounded-noise arm,
$\bar d\le5\times10^{-4}+5\times10^{-3}
=5.5\times10^{-3}\le\mu_\star r=0.033$.
Theorem~\ref{thm:disturbance} gives
$\bar d/\mu_\star\le0.01$. All 5000 safe-projection trajectories satisfy the
finite-time bound and empirically enter this radius by the third update.
Gaussian results remain empirical because the disturbances are unbounded.
\section{Controlled Real-Image Evidence of Contraction Loss}
\label{sec:real_image}

\begin{figure}[t]
\centering
\includegraphics[width=0.8\linewidth]{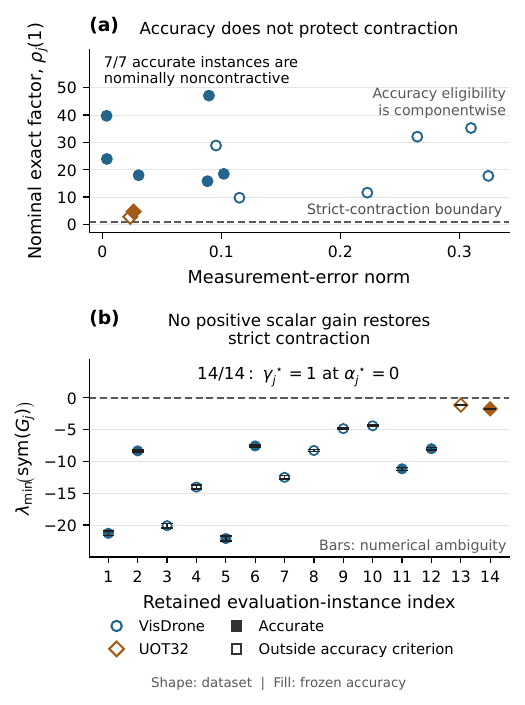}
\caption{Accurate real-image measurements can exhibit scalar-unrecoverable
contraction loss. (a) All seven accurate instances have computed nominal
contraction factors above one. (b) All reported numerical ambiguity intervals
remain below zero. Applying the exact obstruction to the computed segment
operators gives $\gamma_j^\star=1$ at $\alpha_j^\star=0$.}
\label{fig:real_image_geometry}
\end{figure}

We isolate the paper's central mechanism using pretrained detectors and real
images from VisDrone \cite{VisDrone} and the UOT benchmark family
\cite{Panetta2021UOT}, specifically UOT32 \cite{UOT}. Controlled state
perturbations make the state-to-image map and composite Jacobian directly
evaluable. The experiment tests whether accurate measurement values can
coexist with Jacobian geometry that produces nominal noncontraction and
precludes strict contraction over the admissible scalar-gain family. It does not evaluate end-to-end tracking improvement.

\subsection{Controlled Composite Measurement Map}
For each image--branch--state-pair instance $j$, define the composite
state-to-measurement map
\begin{equation}
h_j(x)
=
\operatorname{box\_to\_state}
\left(
p_\phi\left(g_j(x)\right)
\right),
\label{eq:real_composite_map}
\end{equation}
where $g_j(x)$ applies the translation and scale state $x$ to the image,
$p_\phi$ is a fixed pretrained YOLOv8 detector
\cite{Jocher2023UltralyticsYOLO}, and
$\operatorname{box\_to\_state}$ converts the selected detector box to
normalized translation and log-scale coordinates. For image dimensions
$W,H$ and annotated pixel box $(l,t,w,h)$, define
\begin{equation}
\begin{aligned}
c_0&=\left(\frac{l+w/2}{W},\frac{t+h/2}{H}\right),
\quad
s_0=\left(\frac{w}{W},\frac{h}{H}\right),\\
\ell_d&=\frac12\log
\frac{s_{d,x}s_{d,y}}{s_{0,x}s_{0,y}},\\
\tau_d&=c_d-\frac12\bm{1}
-e^{\ell_d}\left(c_0-\frac12\bm{1}\right).
\end{aligned}
\label{eq:box_to_state_coordinates}
\end{equation}
Here $c_d$ and $s_d$ denote the selected detector-box center and size after
removing resize padding and normalizing by the original image dimensions. The
measurement is
$h_j(x)=(\tau_{d,x},\tau_{d,y},\ell_d)^\top$, where $\tau_d$ is normalized
translation and $\ell_d$ is natural-log scale. The state
$x=(\tau_x,\tau_y,\ell)^\top$ lies in
$\mathcal X_{\mathrm{img}}
=[-0.01,0.01]^2\times[-0.02,0.02]$.
We evaluate
\begin{align}
x^+
&=(0.006,-0.004,0.012)^\top,
&\\
\hat x^+
&=(0.004,-0.003,0.009)^\top,
\nonumber\\
x^-&=-x^+,
&
\hat x^-&=-\hat x^+.
\end{align}
Both line segments lie in $\mathcal X_{\mathrm{img}}$. A confidence-eligible
detector branch is selected using ground-truth overlap and fixed along each
segment. Nonmaximum suppression, association, and tracking are excluded.
Set $f(x)=x$, $\widetilde K=I_3$, $\alpha_0=1$,
$\alpha\in[0,2]$, and all added disturbances to zero. The segment-averaged
operators are
\begin{equation}
F_j=I_3,
\qquad
G_j=
\int_0^1
J_{h_j}\!\left(\hat x+s(x-\hat x)\right)\,ds.
\label{eq:real_averaged_jacobian}
\end{equation}
The exact operator, induced factor, and scalar-family optimum are
\begin{equation}
\begin{aligned}
A_j(\alpha)&=I_3-\alpha G_j,\\
\rho_j(\alpha)&=\|I_3-\alpha G_j\|_2,\\
\gamma_j^\star&=\min_{\alpha\in[0,2]}\rho_j(\alpha).
\end{aligned}
\label{eq:real_gamma}
\end{equation}
The controlled measurement is $y_j=h_j(x)$ with $v_j=0$. For
$e_j=x-\hat x$, the resulting error update is
\begin{equation}
e_j^+=(I_3-\alpha G_j)e_j.
\label{eq:real_controlled_update}
\end{equation} The Jacobian $J_{h_j}$ is taken through the complete composite map, including
the image transformation and fixed detector branch. The evaluation requires
no training-data access, model modification, or retraining.

\subsection{Evaluation Protocol and Numerical Validation}

Prespecified criteria retain seven targets from six images and six source
groups, yielding 14 state-pair instances: 12 from VisDrone and two from UOT32.
For the transformed annotation, $y_j^{\rm gt}=x$. Define
\begin{equation}
\begin{aligned}
\varepsilon_j^{\rm meas}&:=h_j(x)-y_j^{\rm gt},
&m_j&:=\|\varepsilon_j^{\rm meas}\|_2,\\
\max_{i=1,2}|\varepsilon_{j,i}^{\rm meas}|&\le0.02,
&|\varepsilon_{j,3}^{\rm meas}|&\le0.15.
\end{aligned}
\label{eq:real_accuracy_test}
\end{equation}
Figure~\ref{fig:real_image_geometry}(a) reports $m_j$, but eligibility is
componentwise and is not determined by the plotted norm.
Because activation switches preclude an assumed global $C^1$ model, the
fixed-branch composite map is locally Lipschitz along each evaluated segment
and hence absolutely continuous there. Its almost-everywhere derivative is
integrable along the segment, and the fundamental theorem for absolutely
continuous functions gives
\begin{equation}
h_j(x)-h_j(\hat x)=G_j(x-\hat x).
\label{eq:real_mean_value_check}
\end{equation}
We compute $G_j$ using adaptive paired 8- and 16-point Gauss--Legendre
quadrature, stopping when
$E_{Q,j}\le10^{-3}+0.005\|G_j\|_2$, where $E_{Q,j}$ denotes the paired-rule quadrature error estimate. Independently, we require
\begin{equation}
\|h_j(x)-h_j(\hat x)-G_j(x-\hat x)\|_2
\le10^{-5}+0.005\|h_j(x)-h_j(\hat x)\|_2.
\label{eq:real_residual_gate}
\end{equation}
All 14 segments pass; the largest residual is $5.21\times10^{-5}$. The plotted
eigenvalue half-width is
$b_j=\max\{10^{-7}\max(1,\|G_j\|_2),2E_{Q,j}\}$. These numerical checks
validate the computation rather than establish absolute continuity, global
$C^1$ regularity, or the robust-envelope assumptions.

\subsection{Accurate Measurements with Scalar-Unrecoverable Contraction Loss}
All 14 nominal operators in Fig.~\ref{fig:real_image_geometry} are
noncontractive, with $\rho_j(1)\in[2.823,47.085]$. Seven satisfy the accuracy thresholds, with detected-to-annotated box
intersection-over-union (IoU) between $0.719$ and $0.907$, and remain
nominally noncontractive. For the computed averaged Jacobians,
\[
\lambda_{\min}\!\left(\operatorname{sym}(G_j)\right)
\in[-22.113,-1.162].
\] Every upper endpoint of the reported ambiguity bands remains negative,
although the bands are diagnostic rather than deterministic enclosures. Let
$z_j$ denote a unit eigenvector associated with the computed minimum
eigenvalue.
For every $\alpha>0$,
\begin{align}
\bigl\|(I_3-\alpha G_j)z_j\bigr\|_2^2
={}&1
-2\alpha z_j^\top
  \operatorname{sym}(G_j)z_j
\nonumber\\
&+\alpha^2\bigl\|G_jz_j\bigr\|_2^2
>1.
\label{eq:real_scalar_obstruction}
\end{align}
Applying \eqref{eq:real_scalar_obstruction} to each segment operator gives
$\rho_j(\alpha)>1$ for every $\alpha\in(0,2]$, while $\rho_j(0)=1$.
Thus, every evaluated instance attains $\gamma_j^\star=1$ at
$\alpha_j^\star=0$ and numerically exhibits scalar-unrecoverable contraction
loss. For the seven accuracy-qualified instances, the computed Jacobian
geometry yields nominal noncontraction and scalar-family unrecoverability even
though the measurement values satisfy the accuracy criterion. 
The matrices $G_j$ are numerically full rank, excluding a sampled-nullspace
explanation. Their negative symmetric-part directions identify adverse
correction orientation. At nominal gain, realized one-step amplification
ranges from $1.102$ to $23.107$. Eight next estimates leave the initial
transformation domain. The analysis is limited to computed one-step operators of fixed pre-NMS
branches. It neither characterizes post-NMS tracking nor establishes
trajectory instability or the regularity and deterministic mismatch bounds
required for online certification. The obstruction applies to the fixed
correction direction, Euclidean metric, and scalar-gain family. Matrix-valued
gains, alternative metrics, and redesigned observers remain possible.
\section{Conclusion}
Accurate learned measurements do not guarantee contractive observer error
dynamics. The measurement Jacobian shapes correction magnitude and
predictor--correction orientation in the exact error operator, allowing
nominal noncontraction even when measurement values satisfy standard accuracy
criteria. The analysis separates failure of a conservative certificate from
scalar-recoverable contraction loss and geometry that lies beyond scalar-gain
repair. When recovery is possible, uncertainty-aware envelopes define a safe
projection that changes the nominal gain only when the contraction target and
same-state improvement are certified. Uniform certification yields exponential unforced error decay and explicit
disturbance bounds. Numerical experiments demonstrate certified recovery and
lower estimation error. Controlled real-image experiments provide numerical
evidence of scalar-unrecoverable contraction loss despite accurate
measurements. The framework determines when scalar retuning suffices and when
the correction structure must change.

\bibliographystyle{IEEEtran}
\bibliography{main}

\end{document}